\documentclass[aps,twocolumn,superscriptaddress]{revtex4-1}
\usepackage{amsmath,graphicx,subfigure,color,natbib,cleveref}
\newcommand{\ave}[1]{\left\langle #1 \right\rangle}
\newcommand{\slfrac}[2]{\left.#1\middle/#2\right.}
\newcommand{\lae}{\lower 2pt \hbox{$\, \buildrel {\scriptstyle <}\over {\scriptstyle
\sim}\,$}}
\newcommand{\gae}{\lower 2pt \hbox{$\, \buildrel {\scriptstyle >}\over {\scriptstyle
\sim}\,$}}
\usepackage{subfigure}
\begin{document}
\title{Robustness of Network Measures to Link Errors}

\author{J. Platig}
\email{jplatig@jimmy.harvard.edu}
\affiliation{Institute for Research in Electronics and Applied Physics, University of Maryland, College Park, Maryland 20742, USA}
\affiliation{Metabolism Branch, Center for Cancer Research, National Cancer Institute, National Institutes of Health, Bethesda, MD 20892, USA}
\affiliation{Department of Biostatistics, Harvard School of Public Health, Boston, MA 02115, USA}
\author{E. Ott}
\affiliation{Institute for Research in Electronics and Applied Physics, University of Maryland, College Park, Maryland 20742, USA}
%
\author{M. Girvan}
\affiliation{Institute for Research in Electronics and Applied Physics, University of Maryland, College Park, Maryland 20742, USA}
\date{\today}

\begin{abstract}
In various applications involving complex networks, network measures are employed to assess the relative importance of network nodes. However, the robustness of such measures in the presence of link inaccuracies has not been well characterized. Here we present two simple stochastic models of false
and missing links and study the effect of link errors on three commonly used node centrality measures: degree centrality, betweenness
centrality, and dynamical importance. We perform numerical simulations to assess robustness of these three centrality measures.  We also develop an analytical theory, which we compare with our simulations, obtaining very good agreement.
\date{today}
\begin{description}
\item[PACS numbers]
89.75.Hc, 87.10.-e
\end{description}
\end{abstract}
\maketitle

\section{Introduction}
\label{sec:intro}
As applications of network science continue to grow and the cost of large data sets decreases, complex network models are increasingly moving from a useful means for building insights \cite{strogatz2001exploring} to a powerful tool for control and prediction \cite{barabasicontrol,schadt2009network}.  However, the false and missing links that often plague these data sets may pose a challenge to the application of complex network models.  For example, networks based on mobile phone records \cite{Onnela01052007} may miss important highly connected hubs due to a lack of institutional phone numbers, while social media-based networks may show friendships between people where no face-to-face friendship exists.  Thus, characterizing the reliability of network properties inferred from measured data with link errors can be an important issue.  This is a challenging problem as there is often no ``true" network to compare against and only an estimate of the link errors can be made. 

Biological networks, in particular, are often constructed from noisy data.  For example, recent high-throughput technologies such as yeast two-hybrid screening now make it possible to test potential interactions between proteins in a organism; however, depending the stringency of the screening, the number of reported protein-protein interactions can vary dramatically \cite{vidalprotein}.  When the number of reported interactions is on the high end, many false links are likely to be included, and when the number of reported interactions is on the low end, many true links are likely missed.   Further, interactions can also be missed when they are conditioned on other events in the cell.  Link errors are also common in the reconstruction of gene regulatory networks from gene expression microarray data; in particular, false links are frequently inferred from non-causal correlations \cite{CLR2007, wailim}.  

While much attention has been devoted to improving network reconstruction algorithms to limit the number of false and missing links \cite{GuimeraSP, CLR2007, basso2005reverse, banavar}, in this paper we aim to provide a step toward understanding the effect of these link errors on the conclusions we draw from network analysis.  
\section{Approach}
\label{sec:appr}

We study the effects of false and missing links on three different network measures of node importance: degree centrality, betweenness centrality, and dynamical importance, which are described below.  In general, our goal is to understand the extent to which a node importance measure calculated using a noisy network correlates with its value in the true network.  In particular, we wish to determine how measures of node importance differ in their robustness to false and missing links.  In this section we describe the different measures of node importance considered, the different types of ``truth" networks studied, and the different models for false and missing links employed.  We limit our considerations to unweighted,  undirected networks with no self-links.

\subsection{Centrality Measures}
The number of links connected to a node is its degree, the most basic centrality measure.  The use of degree has been especially popular in identifying the function of genes in genetic regulatory networks.  Genes with many links can play important roles in multiple biological functions \cite{gersteinreview}. In social networks a node's number of acquaintances or friends reflects the local influence of that node. 

More global measures of node centrality account for a node's neighbors, neighbors of neighbors, and so on.  \emph{Betweenness centrality} is such a measure. The betweenness centrality of a node \emph{i} is defined as \cite{Wasserman1994}
\begin{equation} g(i) = \sum_{j \ne l}\frac{\sigma_{jl}(i)}{\sigma_{jl}},
\end{equation}
where $\sigma_{jl}(i)$ is the number of shortest paths between nodes \emph{j} and \emph{l} going through \emph{i}, and $\sigma_{jl}$ is the total number of shortest paths between \emph{j} and \emph{l}. By summing over all pairs \emph{j,l} we have the fraction of shortest paths that run through \emph{i}.

One might consider a centrality measure that effectively takes into account all paths instead of only shortest paths. There are multiple eigenvalue metrics \cite{perron} that account for such paths.  We focus on the \emph{dynamical importance}.  The dynamical importance of node \emph{i} is defined in terms of the decrease, $-\Delta \lambda_{i}$, of the largest eigenvalue, $\lambda$, of the network's adjacency matrix upon the removal of node \emph{i} \cite{restrepoprl}:
\begin{equation} \label{eq:dit} I_{i} \equiv \frac{\Delta \lambda_{i}}{\lambda}. \end{equation}  
The dynamical importance measure is motivated by the observation that the largest eigenvalue of the adjacency matrix plays an important role in various processes on networks, including synchronization of oscillators \cite{restrepo2006emergence} and phase transitions in boolean models of gene regulatory networks \cite{pompypants}.
\subsection{Model Networks}
\label{sec:model_nets}
In our investigations of the effects of network noise in the form of link errors on the aforementioned centrality measures, we two types of widely-studied networks as our ``truth" networks.  The first type is an Erdos-Renyi (ER) \cite{ergraph} random network.  To construct an ER network with $M$ links we randomly choose $M$ pairs of nodes and draw an edge between each pair.  This kind of network exhibits a Poisson degree distribution if the number of nodes is large.
The other type of truth network we explore is the scale-free (SF) network, which exhibits a power-law degree distribution.  To construct our SF networks, we start with a directed variant of the Barab\'asi-Albert preferential attachment model \cite{barabasialbert}.  Our network begins with a small random seed network to which a single new node is added at every time step.  When each new node is added, two directed links originating from it are made to existing nodes in the network.  These connections are formed such that the probability of linking to an existing node is proportional to its current in-degree.  We then convert this directed network to an undirected network.The resultant network exhibits a degree distribution that is power law in its tail with exponent $\gamma=-2.5$ (in contrast to the $\gamma = -3$ exponent for the original Barab\'asi-Albert construction \cite{barabasialbert}).   
\subsection{Link Error Models}
\label{sec:linker}
In order to explore how link errors affect centrality measures, we consider two models for creating missing and false links.

For both of our link error models, denoted Model 1 and Model 2, we create missing links by randomly selecting $M \delta$ ($0\le\delta\le1$) of the $M$ true links and deleting them.  Models 1 and 2, however, differ in how false links are created.  In Model 1 we create false links by connecting $M \alpha$ node pairs randomly selected from among the $N^2 - N - M$ node pairs not connected by a true link.  From the nodes' point of view, the expected number of its links that get deleted is proportional to its degree in the truth network, while the expected number of links added is independent of its degree.  In our second model of noisy networks (Model 2) , {\em both} the deletion and addition of links occur in proportion to node degree.  Thus in Model 2 false links are added between node pairs where each node in the pair is randomly selected with probability proportional to the node's true degree.  That is, we randomly choose two nodes with probability proportional to their degree; if the two choices do not already have a connecting link, we add a link between them.  We repeat this process until $M \alpha$ links have been added.
\begin{figure}[ht]
\centering
\subfigure{
\includegraphics[width=0.49\textwidth]{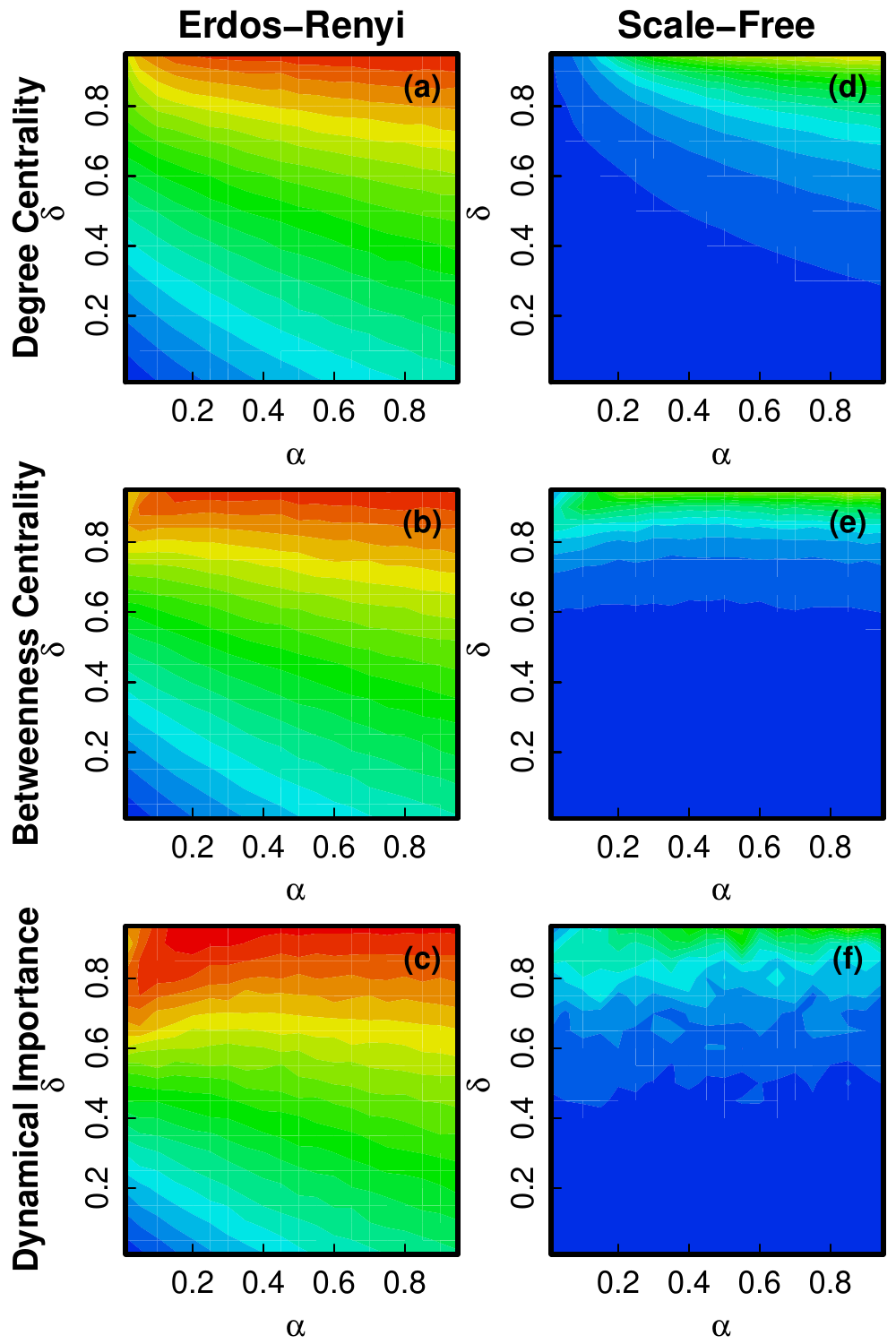}
}
\subfigure{
\includegraphics[trim = 5mm 20mm 5mm 20mm,width=0.25\textwidth]{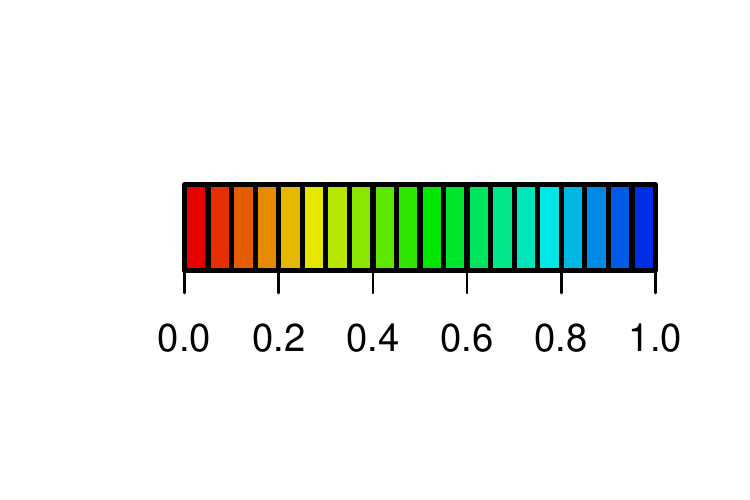}
}
\caption{Model 1: Contour map of correlation before and after introduction of link errors where $\delta$ is the fraction of missing links and $\alpha$ is the fraction of false links for Erdos-Renyi and Scale-Free Networks.  False links are added randomly (Model 1), missing links proportional to the original degree.  True and noisy network measures are perfectly correlated when $\rho$ is 1 (blue) and not correlated when $\rho$ is 0 (red).\label{corpanel1}} 
\end{figure}
\begin{figure}[ht]
\centering
\subfigure{
\includegraphics[width=0.49\textwidth]{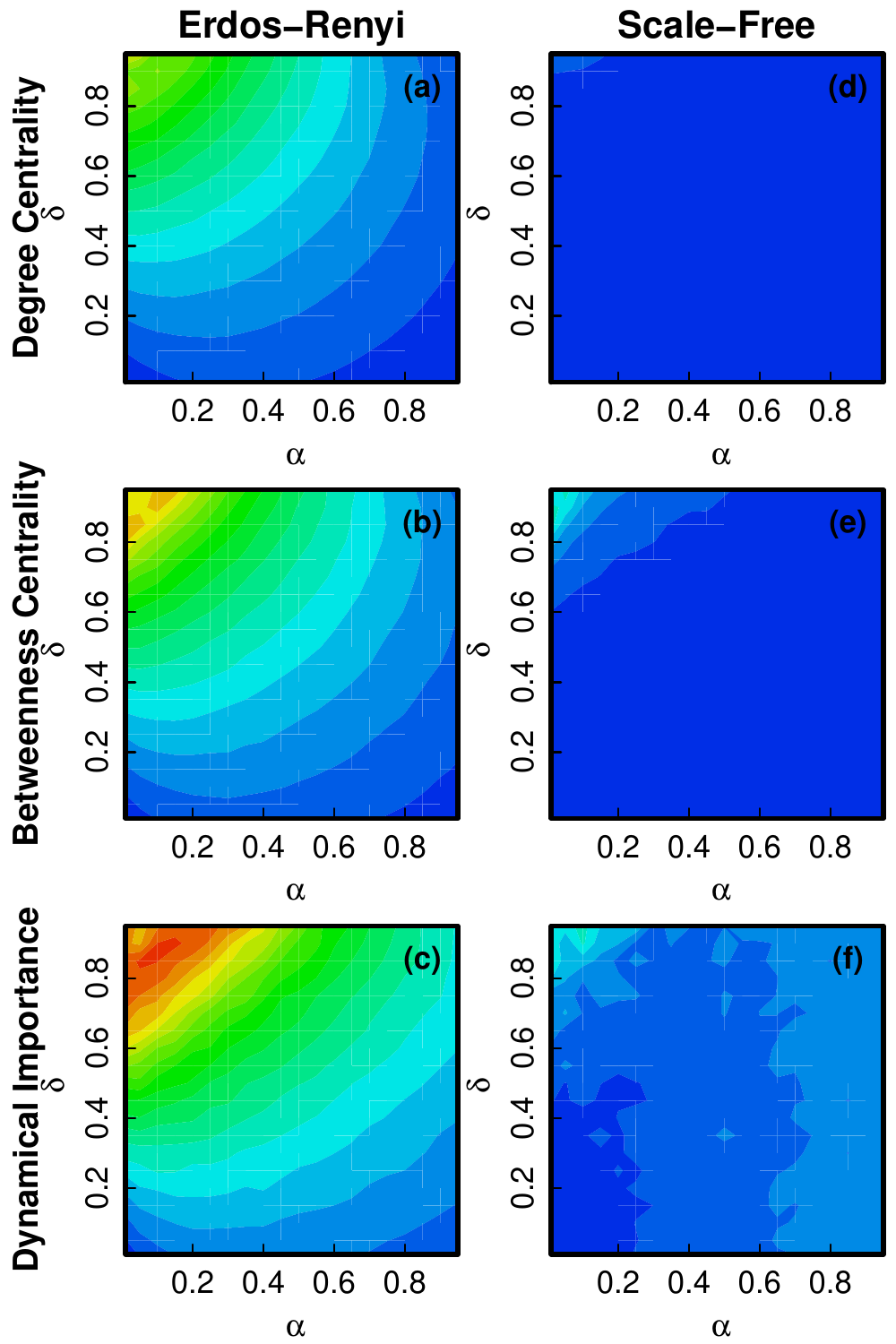}
}
\subfigure{
\includegraphics[trim = 5mm 20mm 5mm 20mm,width=0.25\textwidth]{figaL.pdf}
}
\caption{Model 2: Contour map of correlation before and after introduction of link errors where $\alpha$ is the fraction of missing links and $\alpha$ is the fraction of false links for Erdos-Renyi and Scale-Free Networks.  False and missing links are proportional to the original degree.  True and noisy network measures are perfectly correlated when $\rho$ is 1 (blue) and not correlated when $\rho$ is 0 (red).\label{corpanel}} 
\end{figure}

In what follows we will vary the link deletion fraction $\delta$ and the link addition fraction $\alpha$ to numerically (Sec. \ref{sec:sims}) and analytically (Sec. \ref{sec:anl}) explore the effects of missing and false links.  While $0\le\delta\le1$, note that $\alpha$ can be larger than $1$.  Here we restrict ourselves to $0\le\alpha\le1$ and hence do not consider noisy networks for which the number of false links exceeds the number of true links.  Because some centrality measures are not well-defined when there are multiple disconnected components in a network, only nodes in the giant connected component (GCC) of both the true and noisy network are considered.  

\section{Simulation Results}
\label{sec:sims}
In this section, we report results of numerical simulations investigating the robustness of network centrality measures in the face of link errors for the two different types of truth networks and the two link error models considered.  Using the methods described in Sec. \ref{sec:appr}, we generated Erdos-Renyi networks with with $N=2500$ nodes and average degree $\left<k \right> = 6$ and scale free networks with $2500$ nodes and average degree $\ave{k}=4$.  Starting with each of these truth networks, we then produced noisy variants for different values of $\delta$ (the fraction of true links deleted) and $\alpha$ (the fraction of false links added).  For all values of $\alpha$ and $\delta$, results from the noisy networks were averaged over 25 realizations.  
\subsection{Centrality Correlations}
\label{ssec:simcor}
To assess the effect of link noise on a node's centrality measure $C$, we calculated the Pearson correlation $\rho$ between the true measure $C_{T}$ and noisy measure $C_{N}$ \footnote{We only consider the correlation for each individual centrality measure before and after link errors are added.  We do not study correlations between the different centrality measures, since we regard the latter issue as being more context-dependent.  E.g., it may be more appropriate to choose a centrality measure because its character makes it more indicative of the particular processes that the network is experiencing, than to choose it because it is (by some criterion) more robust.}:
\begin{equation} \label{eq:rho} \rho(C_{T},C_{N}) = \frac{\left<C_{T}C_{N}\right>-\left<C_{T}\right>\left<C_{N}\right>}{\sqrt{(\left<C_{T}^{2}\right>-\left<C_{T}\right>^{2})(\left<C_{N}^{2}\right>-\left<C_{N}\right>^{2})}}, \end{equation} 
where $C$ denotes either the node's degree centrality, betweenness centrality, or dynamical importance, and $\ave{...}$ indicates an average over nodes in the giant connected component of the network.  We used the standard definitions of degree centrality and betweenness centrality from Section \ref{sec:appr} when calculating the correlation $\rho$.  In the case of dynamical importance, we employed a perturbation-based large-$N$ approximation \cite{restrepoprl} of Eq. (\ref{eq:dit}) using the left and right eigenvectors, \emph{u} and $v$, associated with the largest eigenvalue of the network adjacency matrix:
\begin{equation} \label{eq:DI} \hat I_{i} = \frac{v_{i}u_{i}}{v^{T}u}. \end{equation}  
The computational feasibility of Eq. (\ref{eq:DI}) makes it amenable to application in very large networks ($N \approx 50 \, 000$), and it extends naturally to directed networks (only undirected networks are considered in this work).
\begin{figure}[h!]
\includegraphics[trim = 20mm 0mm 0mm 0mm,width=0.48\textwidth]{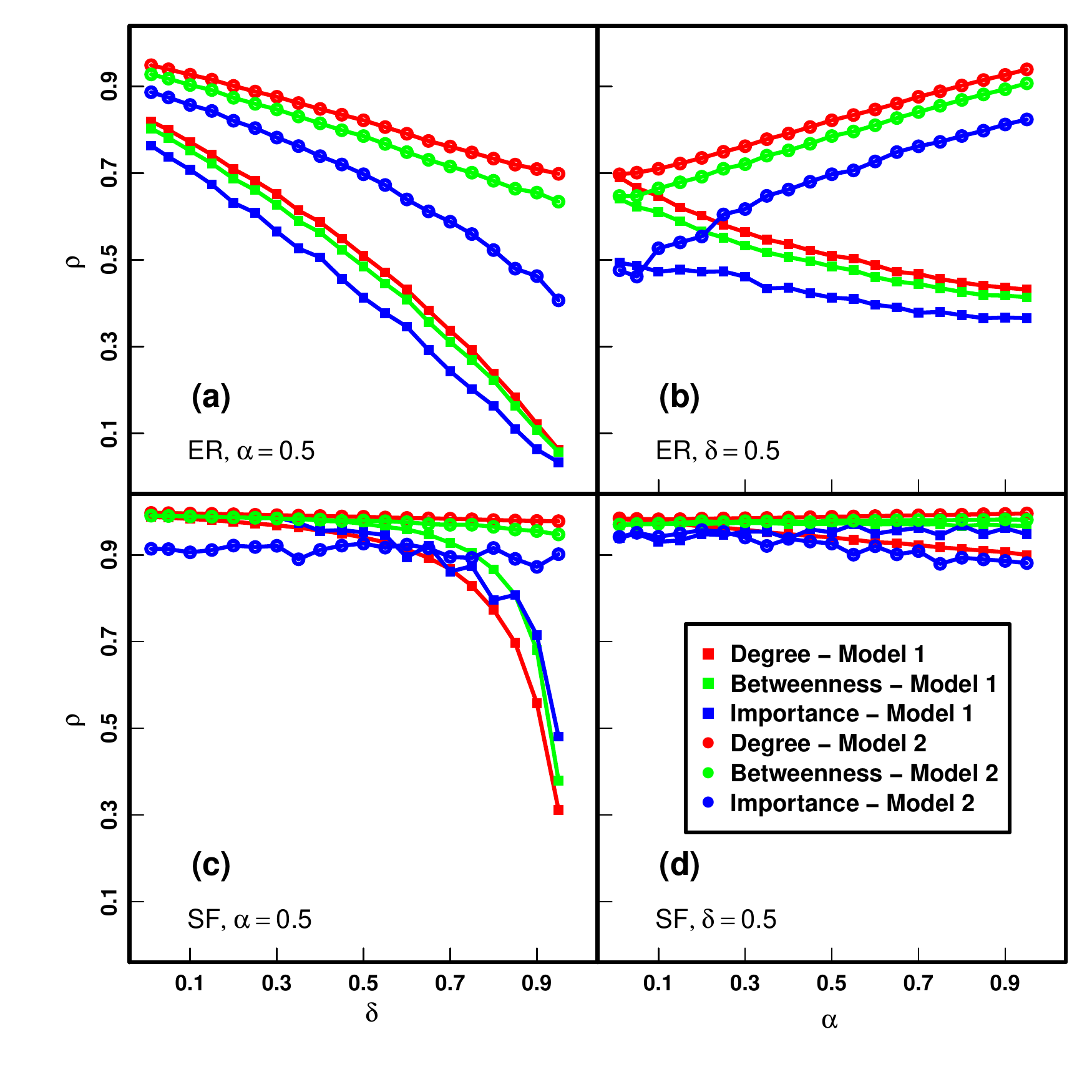}
\caption{The centrality correlation for the three measures studied: degree centrality (red), betweenness centrality (green), and dynamical importance (blue).  Squares correspond to results in which noise is added according to Model 1, circles correspond to results in which noise is added according to Model 2.  Panels (a) and (b) are for Erdos-Reyni truth networks, and panels (c) and (d) are for scale-free truth networks.  In (a) and (c), the fraction of false edges is fixed at $\alpha=0.5$, and the fraction of true edges deleted, $\delta$ is varied.  In (b) and (d), the fraction of true edges deleted is fixed at $\delta=0.5$ and the fraction of false edges added, $\alpha$, is varied. \label{4panel}}
\end{figure}

Simulation results are shown in Figs. \labelcref{corpanel1,corpanel,4panel}.  Figures \ref{corpanel1} and \ref{corpanel} show heat maps of the correlation $\rho$ in $\alpha$, $\delta$-space for the three node centrality measures and the two types of network considered, with Fig. \ref{corpanel1} showing the system behavior for Model 1 link errors and Fig. \ref{corpanel} showing the system behavior for Model 2 link errors.  In order to more quantitatively compare results from the various cases, Fig. \ref{4panel} shows plots of $\rho$ versus $\delta$ with $\alpha$ held fixed at $\alpha=0.5$ (Figs. \ref{4panel}(a) and \ref{4panel}(c)) and of $\rho$ versus $\alpha$ with $\delta$ held fixed at $\delta=0.5$ (Figs. \ref{4panel}(b) and \ref{4panel}(d)).

In Figs. \ref{corpanel1}(a,b,c), which show Model 1 results for our ER network, we see that for low link deletions $\delta \lae 0.5$ all of the centrality measure correlations decrease as $\alpha$ and $\delta$ are increased,  but that this decrease of correlation is somewhat faster when $\delta$ is increases as compared to when $\alpha$ is increased.  At higher false deletion error, $\delta \gae 0.5$, the betweenness, and especially the dynamical importance, become even less sensitive to false link additions ($\alpha$).

Results using Model 1 link errors on SF truth networks (Figs. \ref{corpanel1}(d,e,f)), show that all three centrality measures are significantly more robust to link errors as compared to our results for the ER network, with very small error for values of $\delta \lae 0.7$.  In addition the insensitivity of $\rho$ to $\alpha$ for betweenness centrality and dynamical importance found for the ER network still applies.

Looking at Fig. \ref{corpanel}, which shows results for Model 2 link errors, we again see that the centrality measures for the SF network (Figs. \ref{corpanel}(d,e,f)) are very much more robust to link errors than is the case for the ER network (Figs. \ref{corpanel}(a,b,c)) with perceptible SF error only appearing near $(\alpha,\delta) \approx (0,1)$.  Furthermore, particularly for the ER network, we still see that the correlation between the centrality measures of the true and noisy networks decreases when links are deleted ($\delta$ is increased).  In contrast, with false link additions ($\alpha$ increasing), we find that the correlation actually \emph{increases}.  E.g., for the case of degree (Fig. \ref{corpanel}(a)), this occurs because Model 2 noise is added in proportion to the signal we are measuring (the true degree), and this effect can also be seen for the betweenness centrality and dynamical importance measures (Figs. \ref{corpanel}(b,c)).

Figure \ref{4panel} shows graphs of the correlation $\rho$ along two slices through $\alpha$-$\delta$ space: (i) $\alpha=0.5$ with $\delta$ varying from 0 to 1 (Figs \ref{4panel}(a) and \ref{4panel}(c)), and (ii) $\delta=0.5$ with $\alpha$ varying from 0 to 1 (Figs. \ref{4panel}(b) and \ref{4panel}(c)).  Referring to Figs \ref{4panel}(a), we see that for ER networks at $\alpha=0.5$, as true links are deleted, the correlation decreases more slowly for Model 2 than for Model 1.  As already seen in Fig. \ref{corpanel}(a), Fig. \ref{4panel}(b) shows a pronounced increase of the correlation for ER networks with increase of Model 2 link error additions ($\alpha$) at fixed $\delta=0.5$.  Figure \ref{4panel}(c) shows that for scale-free networks at $\alpha=0.5$ the correlations are relatively insensitive to link deletion for Model 2, while Model 1 shows significant decrease only for relatively large $\delta \gae 0.6$. Finally, we see from Fig. \ref{4panel}(d) that at fixed $\delta =0.5$ the scale-free network is largely unaffected by the addition of false links for both Model 1 and Model 2.

\subsection{Overlap of Highly Ranked Nodes}
\label{ssec:simran}
In addition to correlation robustness, we have also characterized the effect of link errors on the overlap between the top 10\% of nodes in the truth and noisy networks when ranked based on a given centrality measure.  This consideration of overlap is motivated by the fact that node-ranking is often used to select nodes for further study or experimental validation (e.g., see the gene network study of human glioma in Ref. \cite{carro2009}).

\begin{figure}[ht]
\centering
\subfigure{
\includegraphics[width=0.49\textwidth]{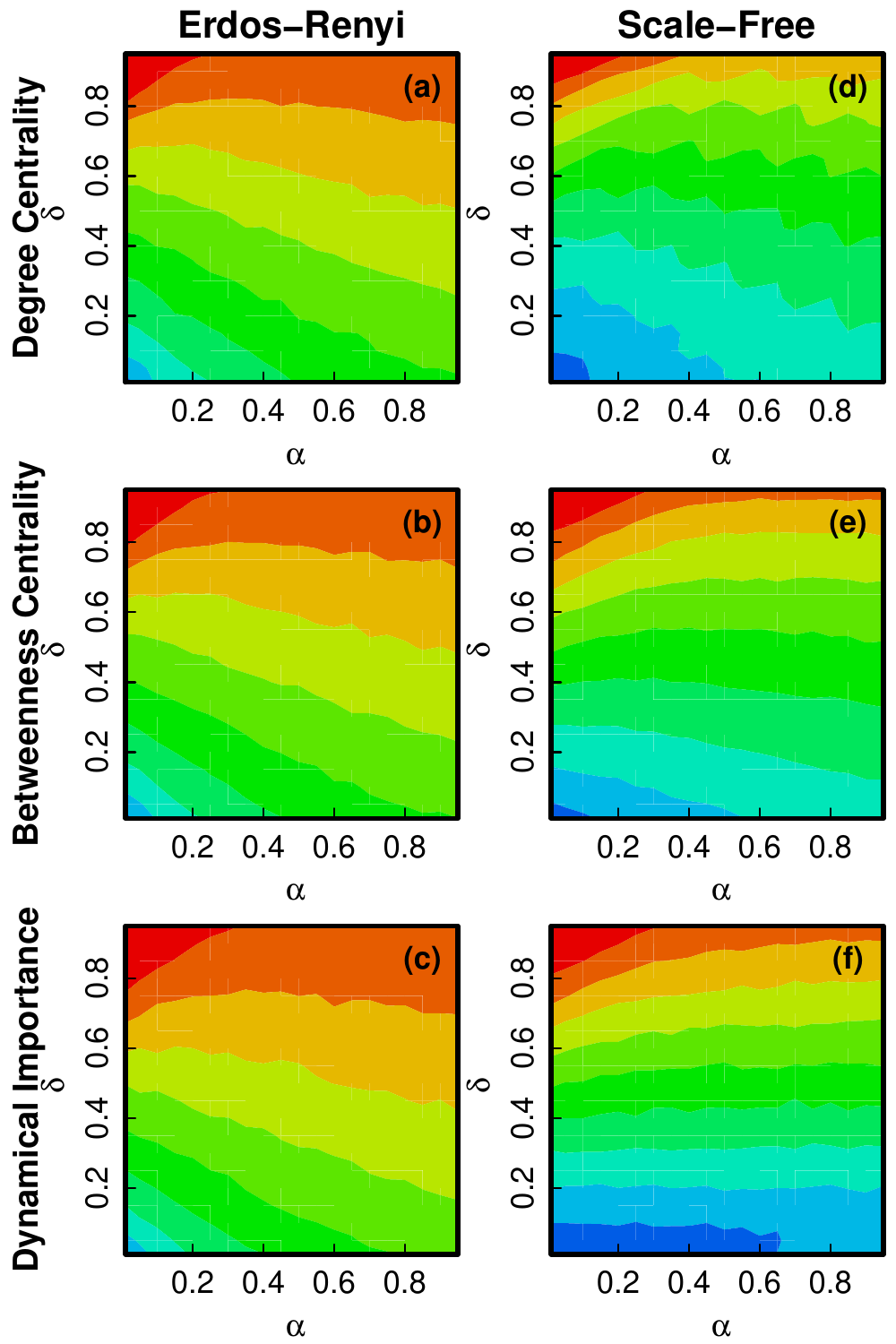}
}
\subfigure{
\includegraphics[trim = 5mm 20mm 5mm 20mm,width=0.25\textwidth]{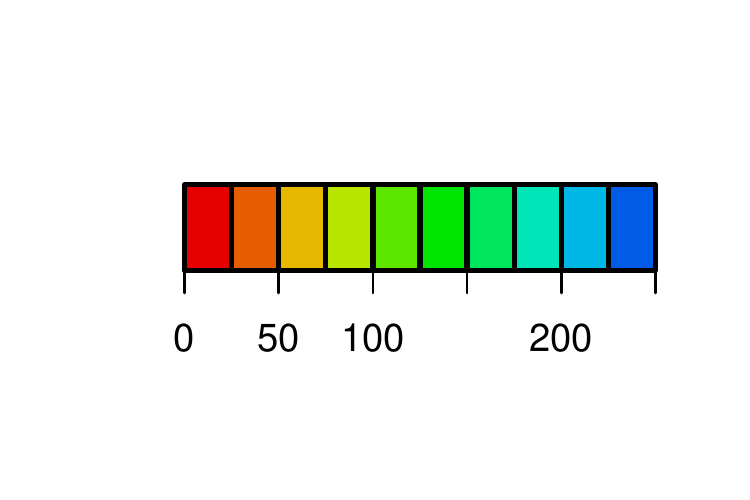}
}
\caption{Model 1: Contour map of the overlap between the top $10 \%$ (250) nodes in the true network and the top $10 \%$ in the noisy network as ranked by each node's centrality measure before and after introduction of link errors where $\alpha$ is the fraction of missing links and $\alpha$ is the fraction of false links for Erdos-Renyi and Scale-Free Networks.  False links are added randomly, missing links proportional to the original degree.\label{overpanel_mod1}} 
\end{figure}

\begin{figure}[ht]
\centering
\subfigure{
\includegraphics[width=0.49\textwidth]{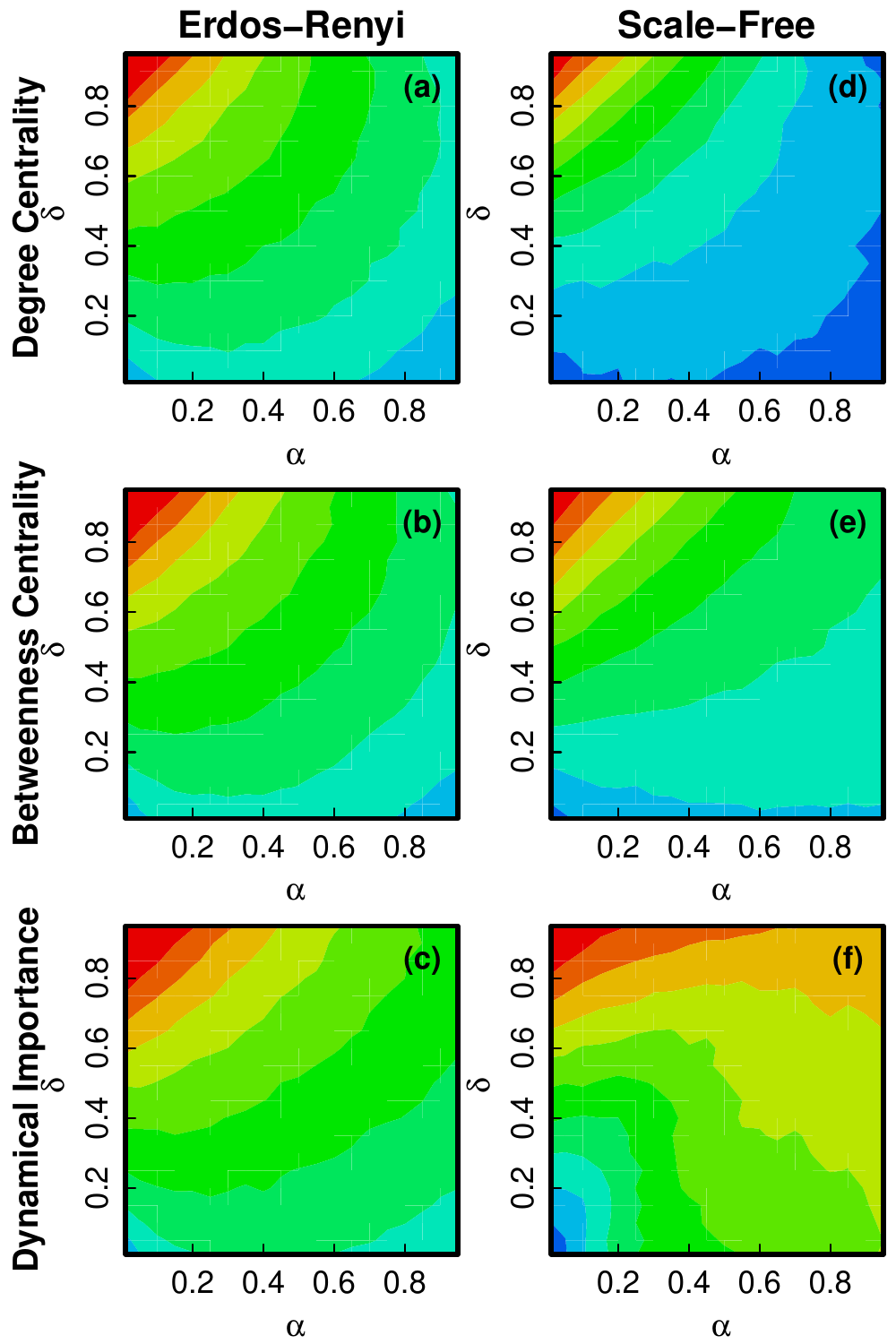}
}
\subfigure{
\includegraphics[trim = 5mm 20mm 5mm 20mm,width=0.25\textwidth]{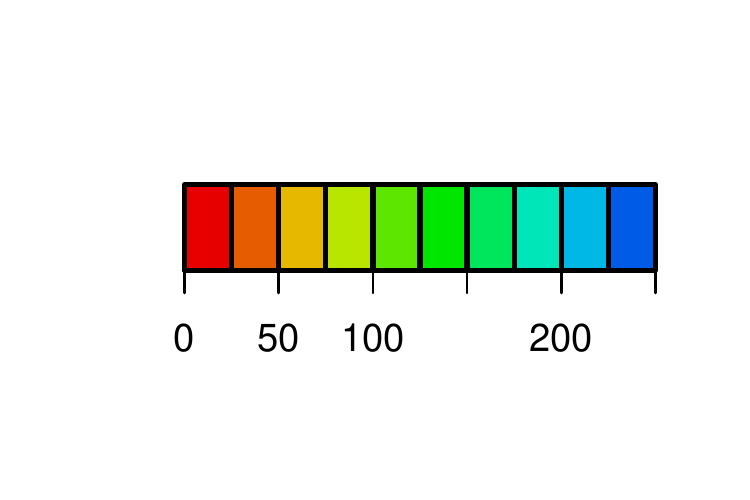}
}
\caption{Model 2: Contour map of the overlap between the top $10\%$ (250) nodes in the true network and the top $10\%$ in the noisy network as ranked by each node's centrality measure before and after introduction of link errors where $\alpha$ is the fraction of missing links and $\alpha$ is the fraction of false links for Erdos-Renyi and Scale-Free Networks.  The number of false and missing links for each node is proportional to the original degree of that node.\label{overpanel_mod2}} 
\end{figure}

With this motivation, we have studied the effects of missing and false links on the ranking of the nodes based on the three centrality measures from Section \ref{sec:appr}.  To do this, we consider the overlap of the top 10$\%$ of the nodes in the giant components of the true and noisy networks and average over 25 network realizations.  Results for Model 1 and Model 2 link errors are shown in Figure \ref{overpanel_mod1} and \ref{overpanel_mod2}.  

While the correlation results, Figs. \labelcref{corpanel1,corpanel,4panel}, show a striking contrast between the ER and the SF networks, with the SF networks being very much more robust to link errors, this result is no longer true when we focus on overlap (Figs. \ref{overpanel_mod1} and \ref{overpanel_mod2}) with the SF and ER networks now \emph{both} showing substantial dependence of the overlap on $\alpha$ and $\delta$.  Similar to the correlation for ER networks with Model 2 link errors, Figs \ref{corpanel}(a,b,c), we now see from Fig. \ref{overpanel_mod2} that the overlap with Model 2 link errors shows substantial decrease with increasing $\delta$, and increase with increasing $\alpha$, applying for both ER and SF networks.

\subsection{Centrality Changes for Individual Nodes}
\label{ssec:nodes}

In Sec. \ref{ssec:simcor} and \ref{ssec:simran} we explored the robustness of the three different centrality measures by assessing the effect of link errors on centrality correlation (Eq. (\ref{eq:rho})) and on overlap of highly ranked nodes, both of which are population-wide characterizations.  In some cases, however, we may be interested in how the centrality of a specific node in the noisy network is related to its centrality in the true network.   In this section, we address such situations.
\begin{figure}[ht]
\includegraphics[trim = 5mm 0mm 0mm 0mm,width=0.49\textwidth]{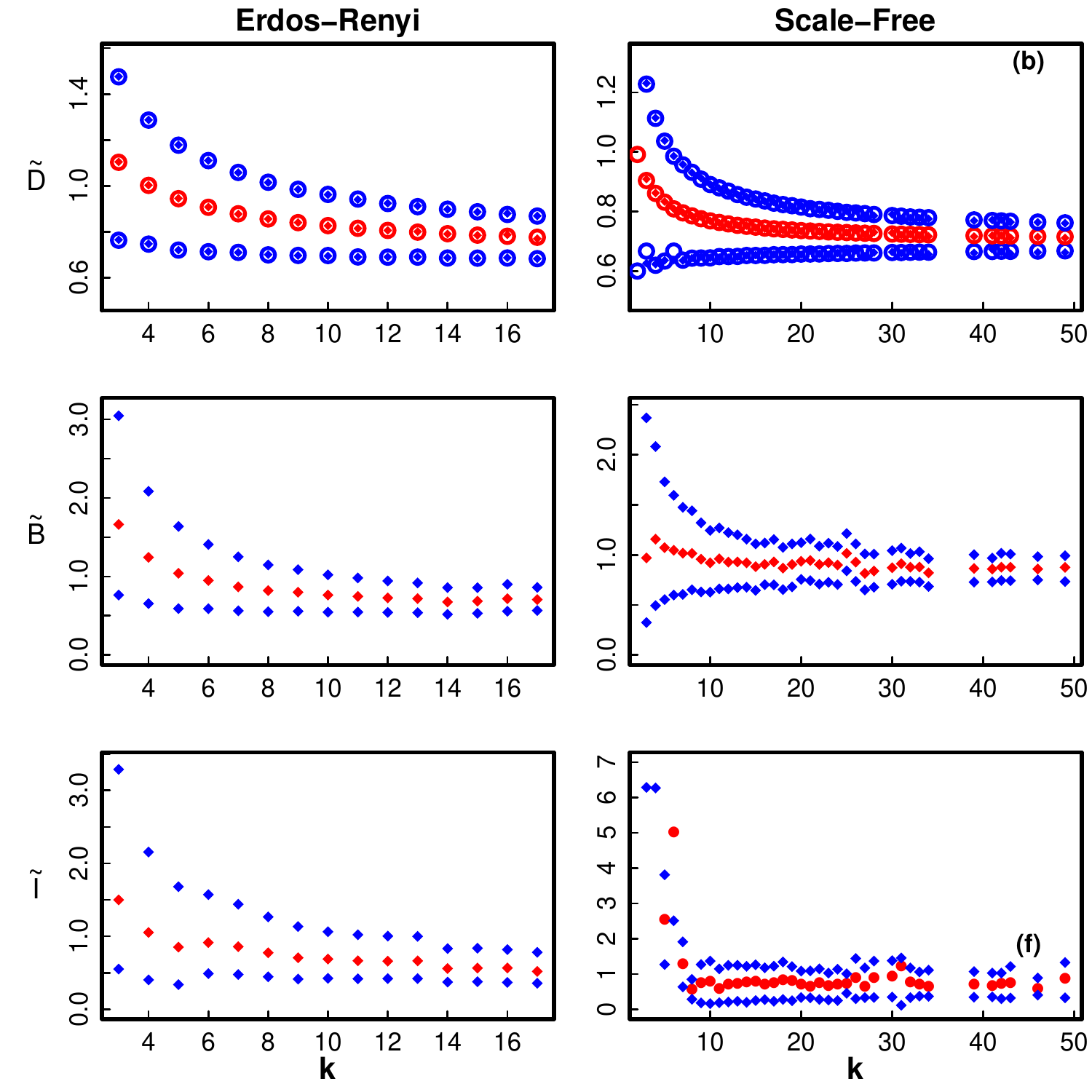}
\caption{Model 1: The first (lower blue), second (red) and third (upper blue) quartiles for the ratio of noisy/true degree ($\tilde{D}$), betweenness ($\tilde{B}$), and dynamical importance ($\tilde{I}$) versus degree (k) in ER (left column) and SF (right column) networks for $\alpha=\delta=0.3$.  The open circles are derived from the theory described in \ref{sec:anl}.   Results are averaged over 500 realizations of the noise model with the same underlying true network.\label{nodepanel1}} 
\end{figure}

\begin{figure}[ht]
\includegraphics[trim = 5mm 0mm 0mm 0mm,width=0.49\textwidth]{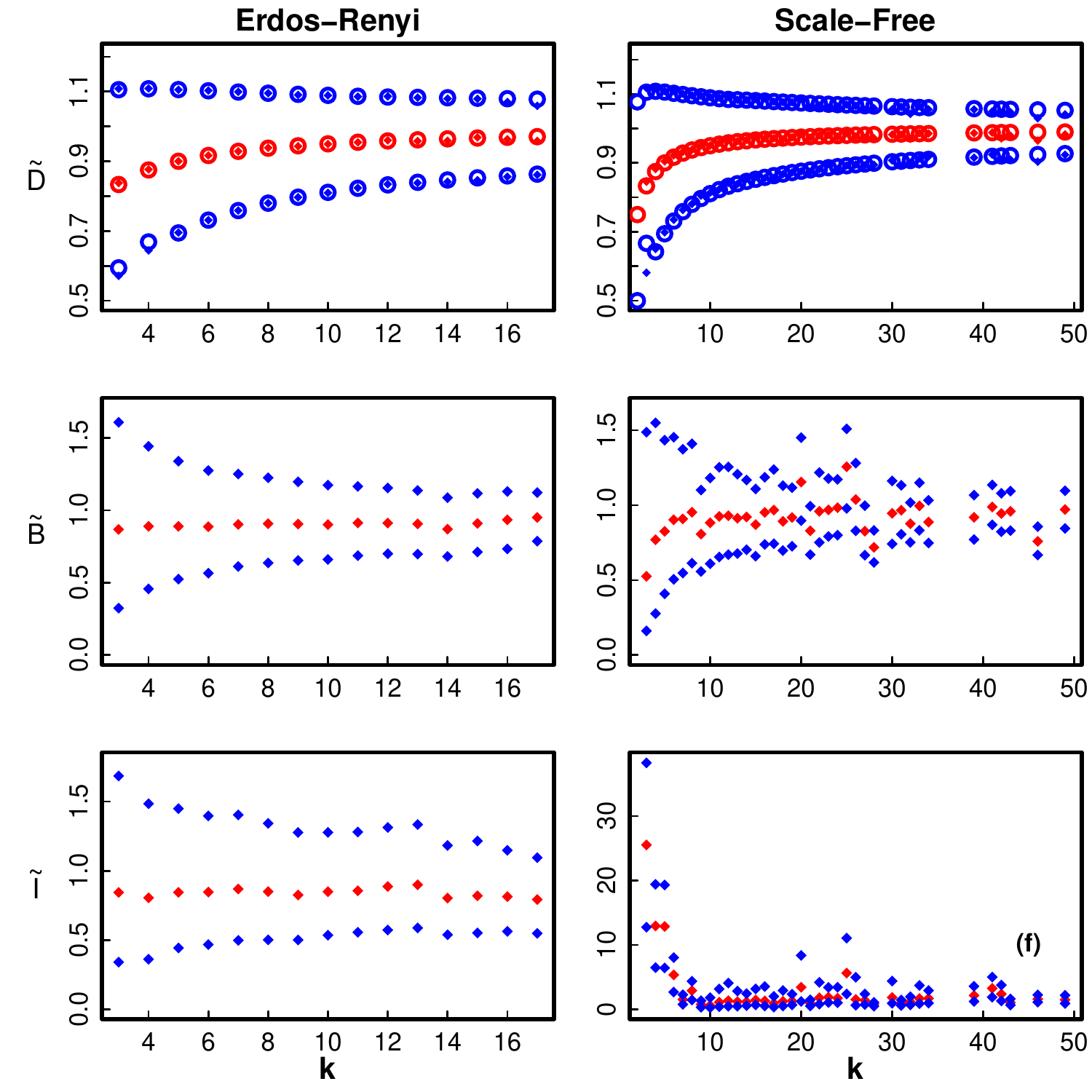}
\caption{Model 2: The first (lower blue), second (red) and third (upper blue) quartiles for the ratio of noisy/true degree ($\tilde{D}$), betweenness ($\tilde{B}$), and dynamical importance ($\tilde{I}$) versus degree (k) in ER (left column) and SF (right column) networks for $\alpha=0.3$, $\delta=0.3$.  The solid curves for the degree are derived from the theory described in \ref{sec:anl}.   Results are averaged over 500 independent realizations of the noise model with the same underlying true network.\label{nodepanel2}} 
\end{figure}

To assess the effects of link errors on the centrality of a specific node with true degree $k$, we consider the set of nodes in the true network with degree $k$, and for each node in that set, take the ratio of its centrality measure (e.g., betweenness) in the noisy network to the same measure in true network.  After repeating the process for 500 realizations of the noisy network (constructed from a single underlying truth network, randomly generated as described in section \ref{sec:model_nets}), we obtain a distribution of the noisy/true centrality ratios for a given value of $k$.  We do this for both ER and SF true networks.  For both models, we focus on an example in which the expected number of false links added is equal to the expected number of true links deleted ($\alpha=\delta=0.3$). The first, second and third quartiles of this distribution are plotted for the three centrality measures with noise generated according to Model 1 in Fig. \ref{nodepanel1} and noise generated according to Model 2 in Fig. \ref{nodepanel2}.   

For Model 1, in which false links are added independent of node degree and true links are removed proportional to node degree, we see that, in both ER and SF networks, the median noisy/true ratio exhibits a general downward trend as $k$ increases.  This trend occurs because low degree nodes are less likely to have links removed than high degree nodes while being equally likely to have links added.   For the betweenness and dynamical importance measures (Figs. \ref{nodepanel1}(c, d, e, and f)), we see that for large degrees the median ratios are very close to 1.  Since the first and third quartile boundaries are also reasonably close to 1 at large $k$, this implies that for higher degree nodes, the betweenness and dynamical importance of a specific node in the noisy network are good predictors of its corresponding measure in the true network, in the case ($\alpha=\delta$) that the total number of links in the noisy network is approximately equal to the number of links in the true network.

For Model 2, in which both link additions and deletions are proportional to the original degree, the median noisy/true ratio approaches one at large degree for all centrality measures, as shown in Fig. \ref{nodepanel2}.  For degree and betweenness centrality, we see that the first and third quartiles are roughly symmetric about the median, whereas for dynamical importance, the third quartile is significantly further from the median than the first, indicating a right-skewed distribution of noisy/true ratios for a given value of degree $k$.  The scatter observed for betweenness and dynamical importance ratios in SF networks (Figs. \ref{nodepanel2}(d and f)) occurs because the noisy networks are built from a single random realization of the true network.  

\section{Analysis of Degree Centrality}
\label{sec:anl}
In this section, we derive analytic approximations for the Pearson correlation $\rho$ of degree centrality before and after the addition of noise and for the probability distribution function of the noisy node degree.  We derive our analytic approximations for both models of link errors studied and show that the predictions of our analytic approximations are consistent with the numerical results presented in the preceding section.
 
As described in Sec. \ref{sec:linker} and employed in our numerical simulations (Sec. \ref{sec:sims}), for both Model 1 and Model 2 we use a micro-canonical procedure in which, for given values of $\delta$ and $\alpha$, the number of falsely deleted links is precisely $M \delta$ (or rather the integer nearest to $M\delta$) and the number of falsely added links $M \alpha$.  However, because this procedure is hard to analyze, to facilitate the theory we employ a closely related canonical procedures that should yield results that are good approximations to the actual Model 1 and Model 2 results.  Specifically, for link deletion, each one of the $M$ true links is deleted with probability $\delta$.  Thus the \emph{average} number of missing links is $M \delta$ with fluctuations whose ratio to the average decreases as $(M \delta)^{-1/2}$, and we expect a good approximation for link deletions when $M \delta \gg 1$.  Similarly for Model 1 link addition, each of the $N^2 - N - M$ pairs of truly unconnected node pairs is connected with probability $\slfrac{M \alpha}{(N^{2}-N-M)}$, creating on average $M \alpha$ false links; while Model 2 addition of truly unconnected node pairs is done with a probability that is proportional to the product of the true degrees of the node pairs.

With the canonical framework assumed \footnote{While the canonical framework is more easily treated by theory, the micro-canonical framework allows faster numerics, and that is why it is used in \ref{sec:sims}}, we first consider the creation of missing links.  If we delete each true link with a probability $\delta$, then the probability that $s$ links are deleted from a node with true degree $k$ is 
\begin{equation} p_{D}(s|k) = \binom{k}{s} \delta ^{s}(1-\delta)^{k-s} . \label{eq:pd}\end{equation}
Next we consider Model 1 link addition. For large $N$, the probability of adding $r$ false links to a node is approximately given by a Poisson distribution,
\begin{equation} p_{A}(r) = \frac{u^{r}}{r!}e^{-u}, \label{eq:pa1}\end{equation}
where $u$ is the average number of false links per node, $u=\slfrac{2M\alpha}{N}$, while for Model 2 the probability that a randomly chosen node has $r$ false links is 
\begin{equation}p_{A}(r|k) = \binom{k}{r}\alpha^{r}(1-\alpha)^{k-r}. \label{eq:pa2}\end{equation}

From our knowledge of $p_{A}$ and $p_{D}$, Eqs. (\ref{eq:pd},\ref{eq:pa1},\ref{eq:pa2}), we obtain the joint probability that a randomly chosen node has true degree $k$ and noisy degree $n$.  Since $n=k-s+r$, 
\begin{align}
p(n|k) &= \sum_{r}p_{A}(r)p_{D}(s=k+r-n|k),\\  
p(n,k) &= p_{0}(k)p(n|k)
\end{align}
where $p_{0}(k)$ is the probability that a randomly chosen node has degree $k$.  In particular for Model 1, 
\begin{equation} \label{eq:pnk1}
p(n|k) = \sum_{r} \frac{u^{r}e^{-u}}{r!} \binom{k}{k+r-n} \delta ^{k+r-n}(1- \delta)^{n-r}, 
\end{equation}
while for Model 2,
\begin{widetext}
\begin{equation} \label{eq:pnk2} p(n|k) = \sum_{r} \binom{k}{r}\alpha^{r}(1-\alpha)^{k-r} \binom{k}{k+r-n} \delta ^{k+r-n}(1- \delta)^{n-r}  \end{equation}
\end{widetext}

\begin{figure}[h!]
\includegraphics[trim = 13mm 0mm 0mm 0mm,width=0.48\textwidth]{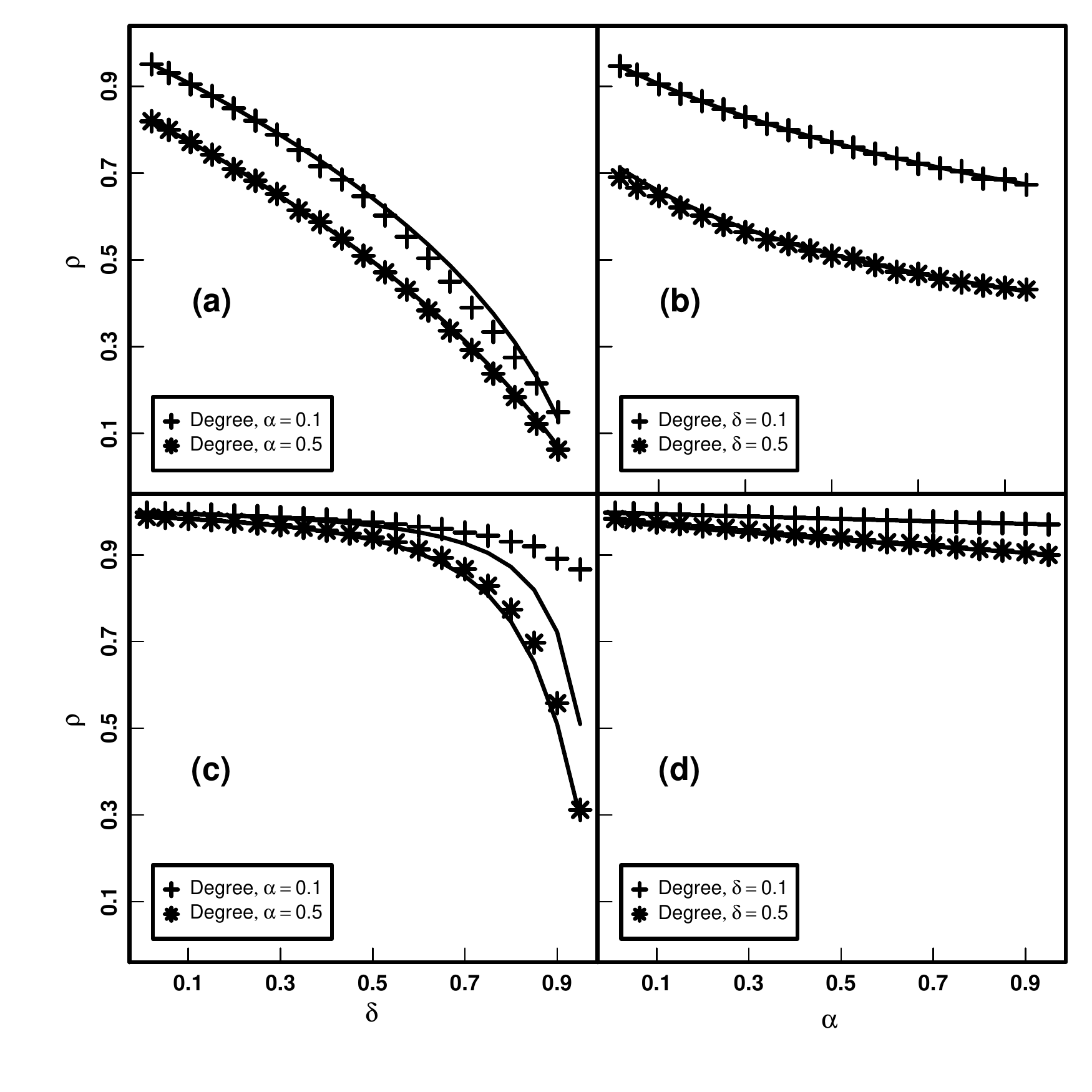}
\caption{Model 1: Pearson correlation between the true degree centrality $k$ and the noisy degree $n$ as a function of missing link fraction, $\delta$, and false link fraction, $\alpha$.  Markers reflect simulation results and theoretical results are plotted as solid lines.\label{fig:theory}}
\end{figure}

The Pearson correlation of ensemble averaged degree, between the true and noisy networks is then:
\begin{equation} \label{eq:rhonk}
\rho( k,n) = \frac{\left<k n\right>-\left<k\right>\left<n\right>}{\sqrt{(\left<k^{2}\right>-\left<k\right>^{2})(\left<n^{2}\right>-\left<n\right>^{2})}}   
\end{equation}
and we use our theory for $p(n,k)$ along with 
\begin{equation}
\ave{k^{x}n^{y}} = \sum_{k,n}k^{x}n^{y}p(n,k),
\end{equation}
to obtain an analytical prediction of $\rho(k,n)$.
\begin{figure}[ht]
\includegraphics[trim = 13mm 0mm 0mm 0mm,width=0.48\textwidth]{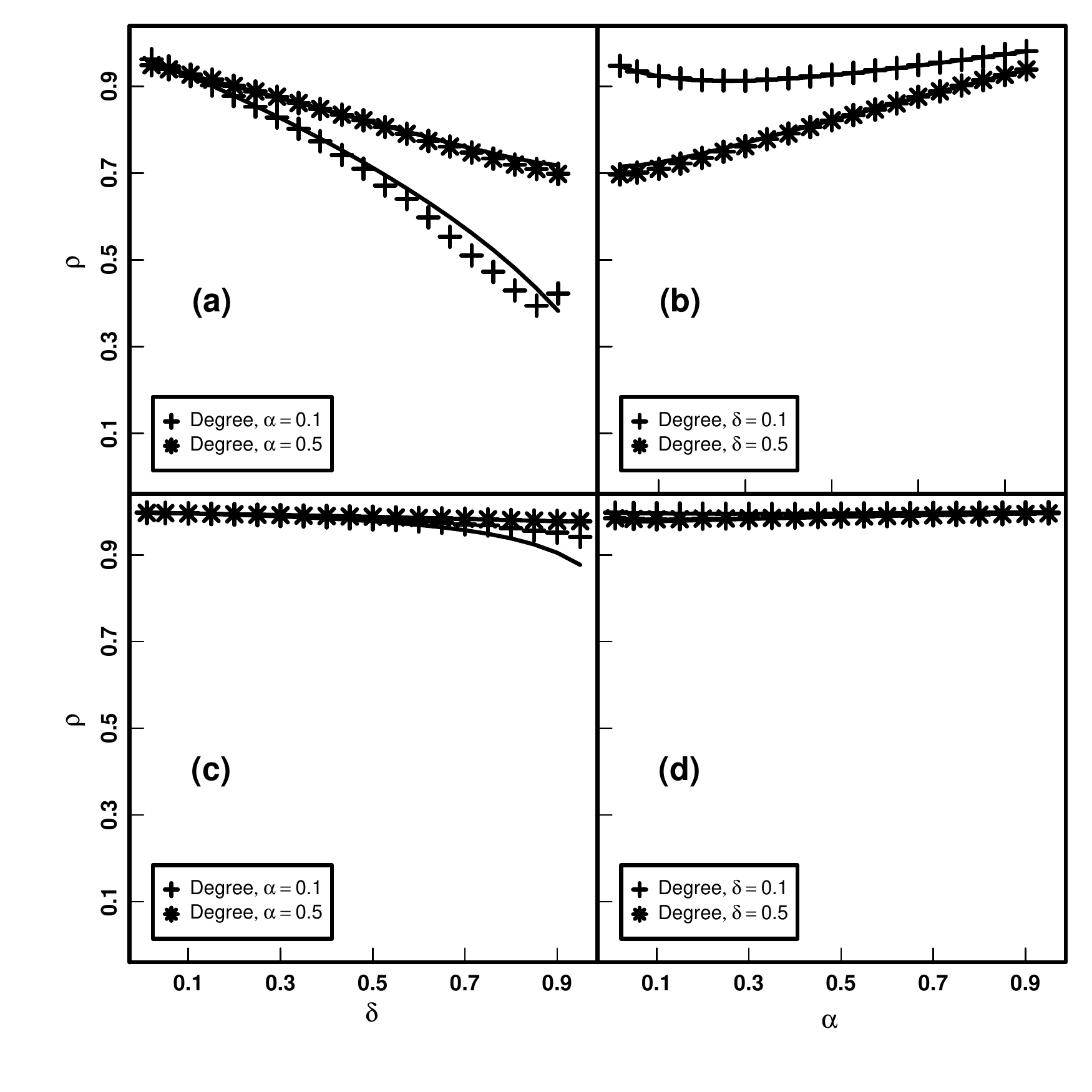}
\caption{Model 2: Pearson correlation between the true degree centrality $k$ and the degree noisy $n$ as a function of missing link fraction, $\delta$, and false link fraction, $\alpha$.  Markers reflect simulation results and theoretical results are plotted as solid lines.\label{fig:theorym2}}
\end{figure}

In order to compare our analysis to our numerical simulations, we take as our $p_0(k)$ the specific numerically generated degree distributions obtained from building our ER and SF networks.  We then use the appropriate forms for $p(n,k)$ from Eq. (\ref{eq:pnk1}) (for Model 1) and Eq. (\ref{eq:pnk2}) (for Model 2) to calculate the expected correlations for degree centrality (Eq. (\ref{eq:rhonk})). 

For Model 1, Fig. \ref{fig:theory} shows comparisons between our numerical simulations (plotted as symbols) and our theory (plotted as lines).  We see that the analytical results are in good agreement with the numerical calculations.  In SF as compared to ER networks, the degree remains strongly correlated in the presence of many false and missing links, with the correlation being driven by the resilient `hub' nodes in the SF networks.  

For Model 2, Fig. \ref{fig:theorym2} shows the theory and simulation of correlation for the degree centrality match well, with a slight discrepancy when $\delta \rightarrow 1$. 

The derived forms of $p(n|k)$ for the two models (Eqs. \ref{eq:pnk1} and \ref{eq:pnk2}) provide theoretical predictions for how the degree centrality $n$ of a specific node in the noisy network relates to its degree centrality $k$ in the true network.  In order to compare the theory to the simulation results discussed in Sec \ref{ssec:nodes}, we used Eqs. (\ref{eq:pnk1}) and (\ref{eq:pnk2}) to find the first, second, and third quartiles of the distribution as a function of $k$, again taking $p_0(k)$ as the specific numerically generated degree distributions obtained from building the ER and SF networks.  Figures \ref{nodepanel1}(a,b) and \ref{nodepanel2}(a,b) show that the theoretical predictions (plotted as open circles) are in very good agreement with the numerical results (plotted as solid diamonds).   
\section{Discussion and Conclusions}

In this paper we have investigated the effect of two types of link errors on three node centrality measures.  We propose two simple models of link error (labeled Model 1 and Model 2) and study their effect for two types of network topology (Erdos-Renyi and scale-free). In Model 1, the probability that a link is deleted depends on the original number of links to which that link is connected, while false links do not depend on the structure of the original network. Model 2 follows the same formulation as Model 1 for deleting links, but in Model 2 the addition of false links is performed with a probability that is proportional to the product of the true degrees of the node pairs. 

We have developed methods for assessing the robustness of node centrality to link errors by comparing the centrality measure for each node before and after link error.  We compare in two ways: (i) by calculating the correlation between the nodes' centrality measures in the true and noisy network (Sec. \ref{ssec:simcor} and \ref{sec:anl}), and (ii) by calculating the overlap between the top $10\%$ of the nodes as determined by their centrality measures in the true and noisy networks (Sec \ref{ssec:simran}).  In the case of correlation we have obtained analytical results (Sec \ref{sec:anl}) which are in good agreement with our numerical simulation results.  The analytical and numerical results for the correlation suggest that degree centrality, betweenness centrality, and dynamical importance are relatively robust to the presence of false edges when the network is scale-free.  Our result for the relative insensitivity of SF networks to $\delta$ is consistent with previous work showing that the size of the giant component in SF networks has a high tolerance to the random removal of nodes \cite{achilles}.  We note, however, that when considering the overlap between top ranked nodes in the true and noisy networks, the much larger SF robustness as compared to the ER case no longer applies.

One common link error not addressed here is that of false edges which complete triangles.  This is a common problem in network reconstruction using microarray data for gene regulatory networks \cite{margolin,wailim}.  In addition to affecting node centrality, these false links may substantially skew the enrichment for network motifs, which are often of interest in biological networks \cite{gerstein2012a,milo2002}.
\section{Acknowledgement}
We thank Shane Squires for his helpful input. This work was supported by the Army Research Office under Grant W911NF-12-1-0101 and by the University of Maryland / National Cancer Institute (NIH) Partnership for Cancer Technology.
\section{References}
%
\end{document}